\documentclass[12pt]{article}
\usepackage{epsfig,amscd,amssymb}

\begin{document}

\newcommand{\nb}{{{\bf n}}}

\title{Haldane-Gapped Spin Chains as Luttinger Liquids:\\
Correlation Functions at Finite Field}
\author{Robert M. Konik and Paul Fendley\\
\smallskip\\
Department of Physics\\
University of Virginia\\
Charlottesville, VA 22904-4714\\}

\date{June 2, 2001}
\maketitle
\begin{abstract}
We study the behavior of Heisenberg, antiferromagnetic, integer-spin chains
in the presence of a magnetic field exceeding the attendant spin gap.
For temperatures much smaller than the gap, the spin chains
exhibit Luttinger liquid behavior.  We compute exactly
both the corresponding
Luttinger parameter and the Fermi velocity
as a function of magnetic field.
This enables the computation of a number of correlators from which we
derive the spin conductance,
the expected form of the dynamic structure factor relevant
to inelastic neutron scattering experiments, and NMR relaxation rates.
We also comment upon the robustness of the magnetically induced gapless
phase both to finite temperature and finite couplings between
neighbouring chains.

\end{abstract}


\newcommand{\del}{\partial}
\newcommand{\nn}{\nonumber}
\newcommand{\gcc}{\Gamma_{cc}}
\newcommand{\la}{\lambda}
\newcommand{\CO}{{\cal O}}
\newcommand{\om}{{\omega}}
\newcommand{\ep}{{\epsilon}}
\newcommand{\lb}{{\langle}}
\newcommand{\rb}{{\rangle}}
\newcommand{\dt}{{d\theta \over 2\pi}}
\newcommand{\dto}{{d\theta_1 \over 2\pi}}
\newcommand{\dtt}{{d\theta_2 \over 2\pi}}
\newcommand{\dttr}{{d\theta_3 \over 2\pi}}
\newcommand{\dtf}{{d\theta_4 \over 2\pi}}
\newcommand{\bd}{{\beta\Delta}}

\section{Introduction}

The existence of a gap in one-dimensional, integer-spin, Heisenberg
antiferromagnets was first predicted by Haldane \cite{haldane}.  He found
that such spin chains can be mapped onto a gapped field theory in 
the large-spin, continuum 
limit.
A variety of checks imply
that this behavior persists down to spin $s=1$.  A spin-$1$ chain
with a specific $(\vec{S}\cdot\vec{S})^2$ coupling has been rigorously
shown to exhibit a spin gap \cite{affleck}.  While at a differing value of the
$(\vec{S}\cdot\vec{S})^2$ coupling, the spin chain 
is gapless \cite{babu}, this critical point is believed to be
unstable in the two-parameter space of couplings.  Gapless behavior
thus only arises as a product of fine-tuning.
Numerous numerical studies carried out
on spin-$1$ chains observe a gap \cite{numerics}.  Experimentally,
inelastic neutron scattering studies on a number of 
quasi-one-dimensional spin-$1$-chain 
materials are consistent with a finite spin
gap \cite{mutka,buyers,renard}.

The physics underlying the gap is particularly robust:
related systems such as two-leg spin-1/2 or Hubbard ladders also
exhibit a gap to spin excitations \cite{dagotto}.  Roughly
speaking, integer-spin composites form across the rungs of the
ladder making it into an effective integer-spin chain.
Both the ability to fabricate these materials 
and their relationship to high $T_c$ cuprate superconductors
have made them the focus of intense theoretical and experimental
study \cite{ladders,so8}.

The centrality of the gap is no less when these materials are
subjected to an applied Zeeman field, $H$.
In a field exceeding some critical value, $H_c \sim \Delta$, 
the excitation spectrum of 
integer-spin chains changes dramatically and the gap 
vanishes \cite{schulz,affleck1}.
At $H=0$ the ground state is a singlet.  The elementary excitations
above the ground state are three massive spin-one bosonic magnon
modes.  In contrast, when $H$ exceeds $H_c$, 
the gap of one of the magnons
closes and the ground state
of the spin chain begins to fill in with gapless excitations.
If interactions between the gapless magnons were completely absent, the
excitations would collapse into a condensate of free bosons.  
On the other hand if interactions were perfectly repulsive at any energy scale,
the bosons could be thought to possess a hard core and so form
a gas of free fermions with some Fermi surface.  But in fact
the interactions are expected to lie midway between these
extremes and what one ends
up with is equivalent to an interacting sea of fermions.  This
of course is nothing more than a Luttinger liquid.  The purpose of this
article is to study its properties.

The field theory describing the continuum limit of integer-spin chains
is the $O(3)$ non-linear sigma model (NLSM) without topological term
\cite{haldane,affrev}. The $O(3)$ NLSM also describes half-integer-spin
chains but in this case the topological term is present with coupling
$\theta = \pi$.  The presence of the topological term leads to gapless
behavior.  Indeed at low energies the $O(3)$ NLSM with $\theta = \pi$
flows to the critical theory, $SU(2)_1$ (for a review see
\cite{affrev}).

Unlike its ancestral theory, the Heisenberg spin-$1$ chain, 
the $O(3)$ sigma model has the virtue of being integrable \cite{zamo,zamo1}.
While aspects of the physics of integer-spin chains have been 
studied using the integrability of the $O(3)$ 
NLSM\cite{affleck1,affleck2,sorensen,sagi},
spin chains remain incompletely understood.
While thermodynamic
properties of an integrable model are generically accessible,
correlation functions are not.
There do exist a variety of techniques to compute
correlators.  At zero temperature, truncated
`form-factor' computations are able to access exact information of the
low-energy properties of the spectral functions \cite{so8,affleck2}.
These techniques have recently been extended to compute exact {\it low
temperature} expansions of these same quantities \cite{rmk}.  There also
exists an elegant semi-classical approach \cite{damle1,damle2}
predicated upon combining ultra-low-energy information from the
quantum theory with older approaches of computing correlators in
classical systems \cite{jepsen}.  However, all of these techniques
require the energy scales in the problem (in particular, the
temperature, $T$, and the applied field, $H$) to be much smaller than
the gap, $\Delta$.

Fortuitously there exists an alternate approach which allows
the computation of correlation functions when $H>H_c$.
The control over the thermodynamic properties of the model
that integrability affords allows 
the computation of the Luttinger parameter.
While the Luttinger
parameter characterizes the strength of the interactions
between excitations, it is a far more central quantity
in that it provides a near complete description of
the low-energy structure of the theory.  Together with the
knowledge of the Fermi velocity, a quantity also easily accessible
with integrability,
a host of information can be determined, including the computation
of specific correlators and their scaling exponents.
Such techniques have been used to study
the sine-Gordon
model in the presence of a chemical potential exceeding the gap
(i.e.\ the mass of the sine-Gordon solitons) \cite{haldane1,papa}.
Similarly the low-energy structure of doped generic Hubbard ladders/armchair
carbon nanotubes has been studied in \cite{so8} utilizing
this formalism.
The particular derivation of the Luttinger parameter in this
paper is based upon the treatment in \cite{fred}.

The ability to compute the Luttinger parameter together with the Fermi
velocity is predicated upon some generic properties of integrable
models.  Most importantly, the exact eigenfunctions of the model's
fully interacting Hamiltonian are known.  With this knowledge comes a
well-defined notion of `particles' or elementary excitations in the
system.  Ultimately this 
feature is a consequence of the infinite number of
conservation laws possessed by the integrable model.
In particular, particle number is conserved in
any collision and multi-particle $S$ matrix elements factorize into
products of two-particles ones.  An integrable model is a
superior version of a Fermi liquid: 
a particle's lifetime is infinite
regardless of distance from the Fermi surface.

In order to appreciate these features of the $O(3)$ NLSM, we 
provide an overview of the model.  The model is described by the
action,
\begin{equation}\label{ei}
S = {1\over 2g} \int dx dt (\del^\mu \nb \del_\mu \nb) ,
\end{equation}
where $\nb = (n_x,n_y,n_z)$ is a bosonic vector field constrained
to live on the unit sphere, $\nb\cdot\nb =1$.  
This action is arrived at from the Hamiltonian
of the spin chain,
\begin{equation}\label{eii}
H = J\sum_i S_i\cdot S_{i+1} .
\end{equation}
In the continuum, large-$s$, limit, the spin operator, $S_i$, is related
to the field, $\nb$, via
\begin{equation}\label{eiii}
S_i = (-1)^i s n_i + M_i .
\end{equation}
$\nb (x,t)$ is the sub-lattice or N\'eel order
parameter while ${\bf M}$ describes the uniform
(i.e. wavevector $k\sim 0$) magnetization.  {\bf M} is related to $\nb $
via
\begin{equation}\label{eiv}
{\bf M} = {1\over g} \nb \times \del_t \nb,
\end{equation}
and so is given in terms of the momentum conjugate to \nb .

The triplet of bosons which form the
low-energy excitations of the $O(3)$ NLSM 
have a relativistic dispersion relation given by
\begin{equation}\label{ev}
E (p) = (p^2+\Delta^2)^{1/2} .
\end{equation}
Here $\Delta$ is the energy gap or mass of the bosons and is related to the
bare coupling, $g$, via $\Delta \sim J e^{-\pi/g}$.  
We have set the bare spin-wave velocity, $v_s=2Js$ 
(the speed of light in this relativistic system) to be 1.
The dispersion
relation of all three bosons is identical as the model has a global
O(3) symmetry.
We stress that this relativistic invariance is a natural feature of the low
energy structure of the spin chain.  

Our approach to the Luttinger liquid phase
of the Heisenberg spin-1 chain shares some similarities
with others taken in the literature.  The NLSM has
been used previously to study the field-induced gapless phase of
spin-1 chains, albeit in the context of large-$N$ techniques \cite{loss},
where the $O(3)$ NLSM is replaced with its $O(N)$ counterpart.
(Parenthetically, we point out that the large N approximation
of the $O(3)$ NLSM have been criticized by \cite{subir}.
In this work the authors note that the ultra low energy
limit of the scattering in the $O(3)$ NLSM differs
from that predicted by a large N expansion of the exact $O(N)$
S-matrix.)

Beyond large-$N$, the Luttinger liquid phase 
of the spin-1 chain has also been studied in the guise of analyzing 
spin-1/2 ladders.  Bosonization techniques have been used to
study the ladder system in a regime where the legs of the ladders
are weakly coupled\cite{approach2}.
In the opposite limit, 
in a regime where there are strong {\it antiferromagnetic}
correlations across the ladder's rungs, the gapless Luttinger
liquid behaviour
has been explored through mapping
the system onto an effective XXZ spin-1/2 chain \cite{approach1}.
In this latter case, the underlying integrability of the XXZ
spin chain can be brought to bear upon the problem \cite{gia}.
Given the simplicity of the map onto the spin-1/2 chain,
we adapt the map to the case where there are {\it ferromagnetic}
interactions across the rungs of the ladder (appropriate
for describing a spin-1 chain).  Our primary purpose in
doing so is to come up with a more precise identification
between the relevant fields in the theory and the bosonic
degrees of freedom of the Luttinger liquid.  Although
we could exploit the integrability of the spin-1/2 chain
to compute the Luttinger parameter, we do not do so.
Given that the map to the effective spin-1/2 chain
is done through a perturbative expansion where the importance
of the higher order terms in the series is uncertain, we 
view extracting the Luttinger parameter directly from
the $O(3)$ NLSM as more reliable.

Haldane gap spin chains in a magnetic field have also been
studied through a mapping onto an interacting, spinless Bose
gas \cite{Takahashi}.  This map provides a reliable description
of the spin chain at small magnetizations.  At small magnetizations,
the low density of excitations forming the ground state interact
only weakly.  The scaling exponents and thermodynamic properties
of the system are 
then independent of the exact nature of these interactions.
Although not done by the authors of
\cite{Takahashi}, the integrability of the Bose gas, like
the integrability of the $O(3)$ NLSM, could be used to 
determine the various properties of the Luttinger liquid
phase.
However beyond the low magnetization regime, the results
would differ.
It is interesting to note, however, 
that by fine tuning the strength of the
interactions of the Bose gas, 
the analysis of the $O(3)$ NLSM, in particular the computation
of the Luttinger parameter, can be reproduced in large degree.
We comment upon this further in Section 3.

In this article we take the Hamiltonian of the spin chain 
to be in its minimal Heisenberg form and
so ignore (for the most part) the affects of
anisotropies upon the physics.  These can take (at least) two forms.
Easy-axis anisotropies,
\begin{equation}\label{evi}
\Delta H = D_x 
\sum_i (S_{xi})^2 + D_y \sum_i (S_{yi})^2 + D_z \sum_i (S_{zi})^2,
\end{equation}
of varying strengths are often
found in spin-$1$ chain materials.
Additionally, actual spin chain materials never take the form of an isolated
chain.  Rather the chains exist in three dimensional arrays with weak
but non-zero interchain couplings, $J'$.  Thus the chains are at best
quasi-one-dimensional.  With a finite $J'$, there will be some
correspondingly finite N\'{e}el temperature, $T_N$.  Below $T_N$
the physics will be dramatically different than described in this
article.

$CsNiCl_3$ was the first material for which evidence of a Haldane
gap was found \cite{buyers}.  This material suffers from the second
aforementioned anisotropy with a relatively large interchain
coupling, $J'/J \sim .017$.  Consequently N\'{e}el order was observed
to set in at $T\sim 5K$.  A more promising material for the observation
of a Haldane gap was found in $Ni(C_2H_8N_2)_2NO_2ClO_4$ (NENP).
For NENP, the ratio $J'/J \sim 6\times10^{-4}$ is sufficiently
small that 3D N\'{e}el order has not been observed down to
temperatures $\sim 1.2K$\cite{renard}.
However this material is characterized by a large easy-axis anisotropy,
$D_z/J \sim .25; ~D_z/\Delta \sim 5/8$ ($D_x \sim D_y \sim 0$).
Related materials $Ni(C_5H_{14}N_2)_3(PF_6)$ (NDMAP) and 
$Ni(C_5H_{14}N_2)_2N_3(Cl0_4)$ (NDMAZ) share similar
easy-axis anisotropies.  In terms of our analysis,
these latter compounds share the additional unwanted feature of 
field-induced antiferromagnetism
\cite{honda}.  The Luttinger liquid
that results from magnetic fields large
enough to extinguish the Haldane gap leads to quasi-long range
antiferromagnetic correlations.  With a small finite $J'$,
these quasi-long range correlations are promoted to full fledged
long range order.  The corresponding N\'{e}el temperature increases
with applied magnetic fields.  
In a mean field framework, we compute this ordering temperature in
Section 4.
Thus at fixed temperature
we expect only a finite window in the applied magnetic field
in which the Luttinger liquid behavior will persist. 

Perhaps the material best suited to the analysis presented
in this paper is $AgVP_2S_6$.  It has an extremely small
interchain coupling, $J'/J \sim 10^{-5}$ and a similarly
small easy axis anisotropy, $D_z/\Delta \sim 10^{-2}$.  However
it possesses a comparatively large gap, $\Delta \sim 320K$.  As
such, high field measurements $(H>\Delta)$ have yet to be
done on this material (and are unlikely to be done soon), 
opposite to the case of NENP \cite{ajiro}, with a
much smaller average gap, $\Delta \sim 20K$.

The paper is organized as follows.  In Section 2 we develop
a Landau-Ginzburg description of the low-energy effective theory of the
integer-spin chain in a magnetic field exceeding the spin gap.
This effective theory reduces to a Luttinger liquid and so has
two controlling parameters: the Luttinger parameter, $K$, and 
the Fermi velocity, $v_F$.
In Section 3 we show how these parameters can be determined
as a function of the applied field, $H$,
through consistency with the $O(3)$ NLSM.

With this description of the low-energy theory in hand, 
we analyze the behavior of a number of correlators in Section 4.  
Using a Kubo formula together with our knowledge of
the current-current correlators, we compute the spin conductance
and the static susceptibility.  With this latter quantity we have
thus come full circle.  We computed $K$ and $v_F$ based upon thermodynamic
considerations and then in turn computed correlators.  From these same
correlators we then (re)compute (consistently) thermodynamic quantities.
We also study the staggered spin-spin correlators, quantities
which would be probed both in inelastic neutron scattering experiments
near wavevector $k=\pi$ and NMR relaxation rate measurements.

While we have already raised the issue of the effects of interchain
couplings upon the stability of the Luttinger liquid phase, we also
consider in Section 4 the robustness of this phase
to finite temperature.
Although our derivation of the effective low-energy theory is done at
zero temperature, the conformal/scale invariant nature of the theory
allows us to easily determine quantities at $T>0$.  To explore how
large the temperature relative to the gap may become before the
Luttinger picture breaks down, we study the susceptibility at finite
$T$ and $H$.  We do so using a more sophisticated description of the
system, the exact equations (good for arbitrary $T$ and $H$) giving
the system's free energy.

\section{Low-Energy Effective Theory for $H>\Delta$}

In this section we describe the emergent Luttinger liquid behavior
of a spin chain arising in magnetic fields
larger than the gap and temperatures satisfying $T \ll \Delta$.
Following \cite{affleck1a},
we provide a corresponding Landau-Ginzburg 
description.  Although we work with an effective theory,
we will be able to compute the various phenomenological
parameters appearing in it by insisting on consistency
with the $O(3)$ NLSM.  This
will form the topic of Section 3.

The Landau-Ginzburg description is an approximate
field theory of the magnons and their interactions. We represent 
the magnon field as ${\bf m}$.  It is a vector under the $O(3)$ symmetry
akin to original field ${\bf n}$ and shares all of
${\bf n}$'s original discrete symmetries.  However 
${\bf m}$ is not constrained to live on
the unit sphere. The magnons have a gap and a relativistic
dispersion relation.  The simplest
effective Hamiltonian with these characteristics is
\cite{affleck1a},
\begin{eqnarray}\label{evii}
{\cal H} &=& {1\over 2}({\bf\Pi}^2 + (\partial_x {\bf m})^2) + 
{\Delta^2\over 2}|{\bf m}|^2
+ \lambda |{\bf m}|^4 - H\cdot ({\bf m}\times {\bf \Pi}),
\end{eqnarray}
where ${\bf \Pi}$ is the momenta conjugate to ${\bf m}$.
We have added a $|{\bf m}|^4$ term to ensure overall stability.
The corresponding Lagrangian (with $H$ in the $z$ direction) is
\begin{eqnarray}\label{eviii}
{\cal L} &=& {1\over 2}( (\partial_t {\bf m})^2
- (\partial_x {\bf m})^2) + H(m_x\partial_t m_y - m_y\partial_t m_x)  
+ {H^2\over 2}(m_x^2+m_y^2) - {\Delta^2 \over 2} |{\bf m}|^2 -
\lambda |{\bf m}|^4 .
\end{eqnarray}
The last three terms form the effective potential for the model.

When $H < \Delta$, the minimum of the Landau-Ginzburg
potential occurs for ${\bf m}=0$.
But when $H>\Delta$ the minima of the potential now occur
for field configurations of the form
\begin{eqnarray}\label{eix}
m_x^2 + m_y^2 &=& {H^2-\Delta^2 \over 4\lambda}\cr
m_z &=& 0.
\end{eqnarray}
As we are interested in the low-energy behavior of the theory, we
focus upon the low-energy fluctuations about these field configurations. 
Introducing the new fields $m$ and $\Phi$ via
\begin{equation}\label{ex}
(m+m_o) e^{\pm i \Phi} = {m_x \pm im_y \over \sqrt{2}} ,
\end{equation}
with $m^2_o = (H^2-\Delta^2)/(8\lambda)$,
the effective Lagrangian can be rewritten (to quadratic order
in the fields)
\begin{eqnarray}\label{exi}
{\cal L} &=& {1 \over 2}\big( (\partial_t m_z)^2 - (\partial_x m_z)^2\big)
+ \big( (\partial_t m)^2 + m_o^2(\partial_t \Phi)^2\big)
- \big( (\partial_x m)^2 + m_o^2(\partial_x \Phi)^2\big)\cr\cr
&& \hskip .5in
+~ 2H (m^2_o + 2m_o m) \partial_t\Phi - 2 (H^2-\Delta^2) m^2 
- {H^2\over 2} m^2_z.
\end{eqnarray}
The low-energy physics in ${\cal L}$ is governed solely by the field
$\Phi$.  The $m$ and $m_z$ modes are massive and may be integrated out.
Doing so results in a Lagrangian of the form
\begin{equation}\label{exii}
{\cal L} = {v_FK\over 2\pi} 
\bigg( (\partial_x \Phi)^2 - {1\over v^2_F} (\partial_t \Phi)^2\bigg),
\end{equation}
the bosonic form of the Luttinger model.
Here $K$ and $v_F$ are the effective Luttinger parameter and
Fermi velocity.  These parameters
can be related to the various parameters appearing
in the original Landau-Ginzburg Hamiltonian.  However
this can be done only at the mean-field level; quantum fluctuations
strongly renormalize their values.
The approach we thus take is to access $K$ and $v_F$ directly
from the integrability of the $O(3)$ NLSM.  This, as has been indicated,
is done in Section 3.

To complete this section we identify the fields of the
$O(3)$ NLSM in terms of $\Phi$.
We can factorize the boson, $\Phi$, into
right and left moving pieces each describing the excitations at the
two Fermi points, $\Phi = \phi_R + \phi_L$.
The (Euclidean) propagators of $\phi_{L/R}$ are given in terms
of $K$.  We have
\begin{eqnarray}\label{exiii}
\lb \phi_L (z) \phi_L (w) \rb &=& - {1\over 4K} \log(z-w);\cr\cr
\lb \phi_R (\bar{z}) \phi_R (\bar{w}) \rb &=& - {1\over 4K} 
\log(\bar{z}-\bar{w}),
\end{eqnarray}
where $z/\bar{z} = v_F\tau \pm i x$.
The fields, $n_\pm (x,\tau )$,
creating $S_z = 1$ excitations are given
in terms of the boson, $\Phi$, by 
\begin{equation}\label{exiv}
n^\pm = e^{\pm i\Phi},
\end{equation}
and so are governed by the propagator of the form
\begin{equation}\label{exv}
\lb n_+ (z) n_- (w) \rb = (x^2+v_F^2\tau^2)^{-{1\over 4K}}.
\end{equation}
We thus understand the spectral functions of low-energy excitations
in terms of $K$ when $H > \Delta$.  Moreover we can 
express the z-component of the magnetization operators in terms
of the $U(1)$ boson currents (in Euclidean space).  Fluctuations
in the magnetization density are given by 
\begin{equation}\label{exvi}
M^3_0 \equiv M_z =  -{K\over v_F\pi}\del_\tau \Phi =
{K\over \pi}(\del_{{\bar z}}\phi_R + \del_{z}\phi_L)
\equiv {1\over 2\pi}(j_R + j_L),
\end{equation}
while the corresponding spin current, $j_S$, equals
\begin{equation}\label{exvii}
M^3_1 \equiv {j_S \over v_F} = -{K\over \pi}\del_x\Phi = 
-{i K\over \pi}(\del_{{\bar z}}\phi_R - \del_{z}\phi_L)
\equiv {i\over 2\pi}(j_R - j_L).
\end{equation}
Current-current
correlators then can be deduced from
\begin{equation}\label{exviii}
\lb j_L (z) j_L(w) \rb = -{K\over (z-w)^2},
\label{currcurr}
\end{equation}
an immediate consequence of (\ref{exiii}).

The above identification of the fields is minimal in nature.  It is
based upon both the Landau-Ginsburg analysis together with the
simplest association of the fields in the NLSM with those of the
Luttinger liquid (\ref{exii}).  However it misses subleading
terms.  To identify such terms directly from the Landau-Ginsburg analysis
would seem difficult.  To circumvent this difficulty, we analyze
a spin-1/2 ladder system.   With ferromagnetic interactions across
an individual rung of the ladder we obtain an effective spin 1.  And
with antiferromagnetic interactions along the length
of the ladder, we obtain an effective antiferromagnetic Heisenberg
spin-1 chain.
In a magnetic field, the ladder system can be reduced
to an equivalent spin-1/2 chain.  Bosonizing such a chain
in the standard fashion
yields a Luttinger liquid, precisely as in (\ref{exii}).
Proceeding in this fashion offers us the advantage that it
allows a more complete identification of the fields, including subleading
terms.  We so find
\begin{eqnarray}\label{exix}
M_z (x,\tau ) &=& 
-{K\over v_F\pi}\del_\tau \Phi (x,\tau ) + c 
\cos (2\pi (\Theta (x,\tau)-Mx));\cr\cr
n^-(x,\tau ) 
&=& e^{-i\Phi (x,\tau )}(1+c_1 \cos(2\pi (\Theta (x,\tau )-Mx))  .
\end{eqnarray}
Here $\Theta (x,\tau)$ is the boson dual to that of $\Phi$ and may
be written as
$$
\Theta (x,\tau ) = {K\over \pi}(\phi_L(z) - \phi_R (\bar{z})),
$$
while $c$ and $c_1$ are some numerical constants.  We see that
the leading terms agree with that of the Landau-Ginsburg analysis.
The subleading terms are interesting in that they depend upon
the magnetization, $M$, of the system, and so lead to incommensurate
behaviour.  This incommensurate behaviour has been argued to 
occur \cite{approach2} in the context of spin-1/2 
ladders with ferromagnetic rungs.

The map from spin-1/2 ladders to effective spin-1/2 chains has been studied
extensively.  Two general approaches have been employed.  In
the case where the spin-1/2 ladder is strongly antiferromagnetically
coupled across the rung 
(we are however interested in ferromagnetic couplings), 
a straightforward map takes the ladder in
a magnetic field to a spin-1/2 chain \cite{approach1,gia}.  
In the case where the spin-1/2
ladder is weakly coupled across the rungs (either ferromagnetically or
antiferromagnetically) more detailed 
studies have been carried out, using
successive bosonizations/fermionizations \cite{approach2}.  
Due to the simplicity of the former, we adapt
this method to ferromagnetic rung couplings, as discussed in Appendix C.

The need to adapt the map is predicated upon two pieces of physics.
The Luttinger parameter of the magnetically induced gapless phase of 
spin-1/2 ladders with ferromagnetic/antiferromagnetic rungs satisfies
$K>1/K<1$.  However a na\"{\i}ve application of the map yields $K<1$ regardless
of the type of rung coupling.
Additionally, the map as it stands does not take into account the different
types of gapless behaviour.  The perpendicular susceptibility
of ladders with antiferromagnetic rungs, for example,
sees power law like decay 
at $k=(\pi,\pi)$ and exponential decay at $k=(\pi ,0)$.
But the situation is reversed for
ladders with ferromagnetic rungs.

\section{Computing $K$ and $v_F$ from the 
Integrability of the $O(3)$ NLSM}

In this section we present exact results for the $H>H_c$ phase
of the $O(3)$ NLSM. They confirm the Luttinger behavior
implied by the Landau-Ginzburg description discussed in the last section.
In particular we show how to extract the phenomenological parameters,
$K$ and $v_F$, appearing in this description
from the integrability of the $O(3)$ NLSM.

We begin by describing how the ground state of the $O(3)$ NLSM
is altered in the presence of a magnetic field.
Neglecting interactions, the
energy of an excitation with bare energy, $E$, in a finite magnetic
field, $H$, along the $z$ axis is
$$
E - S_z H .
$$
This leads to a splitting of the spin-1 triplet of magnons.  In a
mean-field `semi-conductor' picture, this splitting is sketched in
Figure 1.  As the magnetic field, $H$, is increased beyond the critical
value of $H=H_c=\Delta$, the ground state of the spin chain
dramatically changes.
The ground state
now begins to fill in with a finite density of magnons (and so
is akin to a doped semiconductor).
If the $S_z = 1$ magnons were non-interacting,
they would condense in the lowest possible energy level.  But the
magnons interact repulsively, and so they fill the ground state as if
they were fermions or hard-core bosons.
The technical manifestation of
this behavior is the minus sign in the $S$-matrix at zero momentum
transfer ($\theta=0$ in (\ref{exx})).
With $H$ larger than $\Delta$, the energy of an excitation
is potentially negative and 
excitations carrying $S_z=1$ appear in the ground state.  We
illustrate this in Figure 1 schematically
by plotting the dispersion relations for
the three types of excitations in the system: $S_z = -1,0,1$.  
\begin{figure}
\centerline{\epsfxsize=5.0in\epsffile{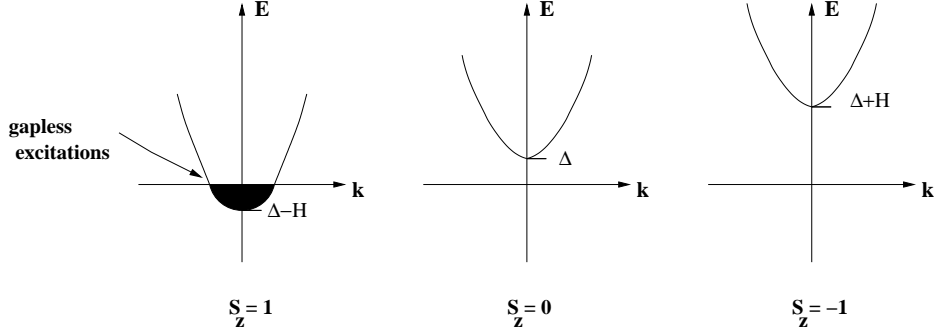}}
\caption{A sketch of the ground state of the $O(3)$ non-linear
sigma model in
the presence of a magnetic field that exceeds the gap, $\Delta$.
The $S_z=1$ band is now partially filled with excitations;
the $S_z=0$ band is unchanged as it does not couple to the magnetic field;
and the $S_z=-1$ band is shifted upwards in energy.}
\end{figure}

The ground state of the system for $H>H_c$ is therefore a sea of magnons.
The low-energy spin excitations consist
of slight deformations of the filled sea in the $S_z=1$ band. 
These low-lying excitations
can be of arbitrarily small energy and so are gapless.
If we linearize such excitations
above the Fermi energy of this band, we obtain a Luttinger liquid
characterized by a Luttinger parameter, $K$, and a Fermi velocity, $v_F$.

To describe the ground state exactly we need to take
into account interactions of the particles.
These interactions are parameterized by their scattering.
As the $O(3)$ sigma model is integrable, the scattering matrix is
known exactly.  
Exploiting the theory's
relativistic invariance, we can parameterize a particle's
energy and momentum in terms of its rapidity $\theta$, defined as
$$E=\Delta\cosh(\theta );\qquad\qquad P =\Delta \sinh (\theta ).$$  
In this
parameterization, Lorentz boosts send $\theta \rightarrow \theta
+\alpha$.  Lorentz invariant quantities like $S$ matrix elements
therefore depend only on differences of rapidities.  
The $S$ matrix for scattering the two-particle state
$|a_1(\theta_1 ) a_2(\theta_2)\rangle$ ($a_1,a_2=1,2,3$) into 
the final state $|a_4(\theta_2) a_3(\theta_1)\rangle$
is \cite{zamo}
\begin{eqnarray}\label{exx}
S^{a_3a_4}_{a_1a_2} (\theta_{12} ) = 
\delta_{a_1a_2}\delta_{a_3a_4}\sigma_1 (\theta_{12} ) +
\delta_{a_1a_3}\delta_{a_2a_4}\sigma_2 (\theta_{12} ) +
\delta_{a_1a_4}\delta_{a_2a_3}\sigma_3 (\theta_{12} ),
\end{eqnarray}
where $\theta=\theta_1-\theta_2$ and
\begin{eqnarray}
\nonumber
\sigma_1 (\theta ) &=& 
{2\pi i \theta \over (\theta + i\pi)(\theta - i2\pi)};\cr
\sigma_2 (\theta ) &=& 
{\theta (\theta - i\pi) \over (\theta + i\pi)(\theta - i2\pi)};\cr
\sigma_3 (\theta ) &=& 
{2\pi i (i\pi - \theta) \over (\theta + i\pi)(\theta - i2\pi)}.
\end{eqnarray}
Since the ground state is filled with solely $S_z=1$ particles, we
need the $S$ matrix element, $S_{++}$, for scattering of two such
particles:
\begin{equation}\label{exxi}
S_{++} (\theta) = {\theta -i\pi \over \theta + i\pi}.
\end{equation}
This is found from the change of basis, $A_{\pm} =
{1\over\sqrt{2}}(A_1+iA_2)$, where here $A_{i}$ is an operator
creating an excitation carrying quantum number $i$.

Knowing $S$ allows us
to determine the density of states per unit length, $\rho(\theta)$,
of the sea of particles at
$H>H_c$. 
The derivation is standard and can
be found for the case at hand in \cite{hasen}. 
For completeness we repeat these arguments
in Appendix A.
At zero temperature, the repulsive nature of the 
particles leads them to fill the sea up to some
Fermi momentum, so that the density satisfies $\rho(\theta)=0$ for
$|\theta|>\theta_F$.
For $|\theta|<\theta_F$, we have
\begin{equation}\label{exxii}
\rho(\theta ) = {\Delta \over 2\pi} \cosh (\theta )
+ \int^{\theta_F}_{-\theta_F} d \theta' \rho (\theta ') 
\Gamma_{++} (\theta-\theta'),
\end{equation}
where
\begin{equation}\label{exxiii}
\Gamma_{++}(\theta ) \equiv
{1\over 2\pi i}\partial_\theta \log S_{++}(\theta )
 = {1\over \theta^2 + \pi^2}.
\end{equation}
The first term on the right-hand-side of (\ref{exxii}) is the free term,
while the kernel, $\Gamma_{++}$, appearing in the second term
measures the strength of the
interactions.  As $\Gamma_{++} > 0$, the interactions lead to 
a density of states greater than the bare value of
${\Delta \over 2\pi}\cosh (\theta )$ associated with purely free fermions.
The strength of the magnetic field, $H$, determines the Fermi rapidity,
$\theta_F$. To find it, it is first convenient 
to introduce the dressed energy,
$\epsilon(\theta)$, of the magnons. This is the amount of energy the system
loses when a particle of rapidity, $\theta$, is removed from the sea. 
It is given by
\begin{eqnarray}\label{exxiv}
\ep (\theta ) &=& \Delta \cosh (\theta ) - H + \int^{\theta_F}_{-\theta_F} 
d\theta' \ep(\theta') \Gamma_{++}(\theta - \theta').
\end{eqnarray}
At zero temperature, the particles fill all allowed levels up to 
$\theta=\theta_F$. Therefore the Fermi rapidity is determined by solving
(\ref{exxiv}) subject to the boundary condition,
$$\ep (\theta_F ) = 0. $$ 
Thus for $|\theta|<\theta_F$, $\epsilon(\theta)<0$.
The energy of the system per unit length at zero temperature is given by
\begin{equation}\label{exxv}
E(H) = \int_{-\theta_F}^{\theta_F}
\rho(\theta)\left[\Delta\cosh (\theta )- H\right] =
 \Delta \int_{-\theta_F}^{\theta_F} \frac{d\theta}{2\pi}\, \epsilon(\theta) 
\cosh (\theta ) .
\end{equation}
The zero-temperature
magnetization follows immediately
from the above equation,
\begin{eqnarray}\label{exxvi}
M(H) &=& -\partial_H E(H), 
\label{suscep}
\end{eqnarray}
with the corresponding susceptibility given by $\chi (H) = \partial_H M(H)$.

\begin{figure}
\hskip.7in \rotatebox{-90}{\epsfxsize=3.2in\epsffile{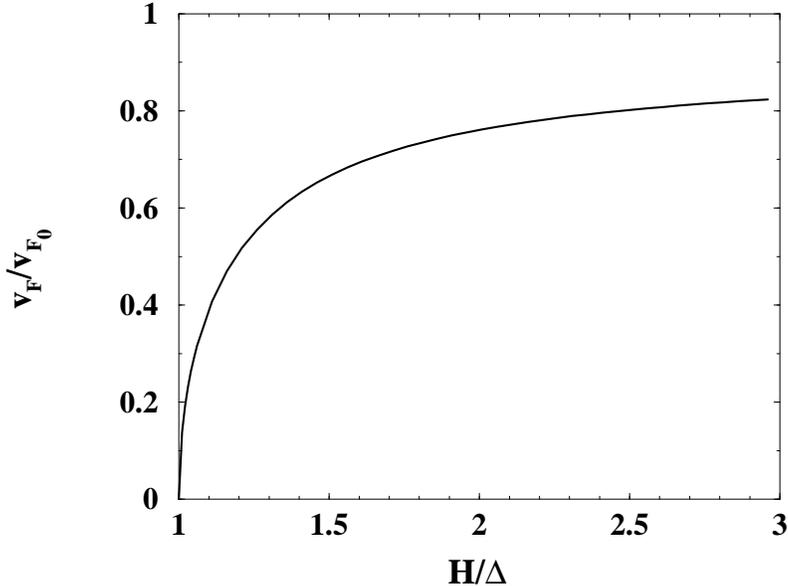}}
\caption{A plot of the Fermi velocity of the low-energy
excitations present in the $S_z=1$ band when $H>\Delta$.}
\end{figure}

We now are able to compute the Luttinger
parameters $K$ and $v_F$ appearing in the effective theory
as functions of $H/\Delta$.  The
key is to study this system in terms of its gapless quasi-particle
excitations present when $H>\Delta$. These
excitations are not the original magnons, but rather the excitations
above the magnon sea.  Since they are gapless, they move either to the
right or left at the Fermi velocity.  We define their rapidity
$\theta$ via
$\ep(\theta ) = \pm v_F p(\theta ) = \mu e^{\pm\theta}$, where the
mass scale $\mu$ is
arbitrary and can be redefined by a shift in $\theta$.  To compute
$v_F$ for the gapless excitations above the magnon sea, we first note that
\begin{equation}\label{exxvii}
v_F = {\del\ep \over \del p}\Bigg|_{\ep=\ep_F} = {\del\theta \over \del p}
{\del \ep \over \del \theta}\Bigg|_{\theta=\theta_F}.
\end{equation}
Here $p$ is the dressed momentum of the excitations and is given
by an equation similar to the one governing the energy:
\begin{equation}\label{exxviii}
p (\theta ) = \Delta\sinh (\theta ) + \int^{\theta_F}_{-\theta_F} 
d\theta' \rho (\theta') \phi(\theta-\theta'),
\end{equation}
where $\phi (\theta ) = \log S_{++} (\theta ) /2\pi i$ is
the unrenormalized scattering phase.
By comparing (\ref{exxviii}) with (\ref{exxii}) 
we see $\del_\theta p = 2\pi\rho$.
The Fermi velocity is then given by
\begin{equation}\label{exxix}
v_F = {1\over 2\pi\rho(\theta ) }
{\del \ep \over \del \theta}\Bigg|_{\theta=\theta_F}.
\end{equation}
While the integral equations (\ref{exxii} and \ref{exxiv}) 
cannot be solved in closed form,
it is easy to solve numerically. This yields the curve
plotted in Figure 2.  Its most notable feature is the
vanishing of $v_F$ for $H$ near $\Delta$.
Below we will give a power-series expansion for $v_F$, valid when
$H$ is near $\Delta$.

\begin{figure}
\hskip.7in \rotatebox{-90}{\epsfxsize=3.2in\epsffile{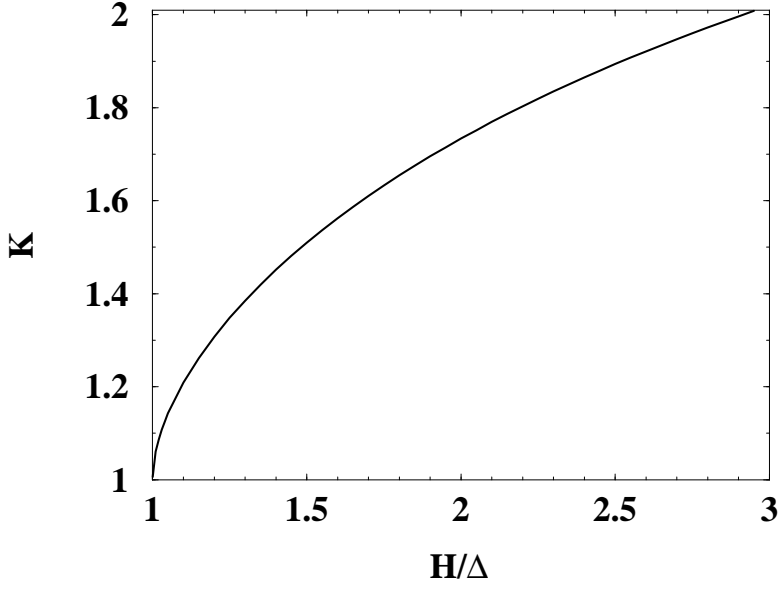}}
\caption{A plot of the Luttinger parameter describing the low
energy excitations present in the $S_z=1$ band when $H>\Delta$.}
\end{figure}

The coupling $K$ is similarly straightforward to find.
As we now show, it is related to the renormalized spin, $S_R$,
of excitations near the Fermi surface.
$S_R$ measures how interactions change the response of a spin $S_z = 1$
excitation near the Fermi surface to a change in the magnetic field.
It is related to the dressed energy by
\begin{eqnarray}\label{exxx}
S_R &=& -\del_H \ep (\theta )\big|_{\theta=\theta_F},
\end{eqnarray}
and so obeys the integral equation,
\begin{eqnarray}\label{exxxi}
\label{eIIIxiv}
S_R &=& 1 + \int^{\theta_F}_{-\theta_F} S_R(\theta ') \Gamma_{++} 
(\theta_F-\theta').
\end{eqnarray}
The $H$ dependence of $S_R$ arises from the dependence
of $\theta_F$ on $H$.

The basic idea behind the relationship of $K$ and $S_R$
is that the low-energy excitations
near $\theta=\theta_F$ are free-fermionic.  That is, their dressed S-matrices
are $-1$ \cite{fred}.  In contrast to \cite{fred}, here it is relatively easy
to demonstrate this and we do so in Appendix B.
As a consequence of the free-fermionic behavior, the only non-vanishing
matrix elements of the current
operator, $j_L = -2K(v_F^{-1}\partial_\tau - i\partial_x )\phi_L$ ,
involve a single particle-hole pair. 
Lorentz invariance requires
that these matrix elements be given by 
\begin{equation}\label{exxxii}
\lb \theta_h\,\theta_p |j_L(0)|0\rb = \mu c e^{(\theta_h+\theta_p)/2},
\end{equation}
where $\theta_{h}$ and $\theta_{p}$ are the rapidities of the particle
and hole respectively.
The constant $c$ is simply related to both $K$ and $S_R$.
In a Luttinger liquid, $K$ appears in the current-current
correlator (\ref{exviii}). To relate this to $c$, we
insert a complete set of states
and use the matrix element (\ref{exxxii}), giving
\begin{eqnarray}\label{exxxiii}
\lb j_L (x,t) j_L (0) \rb &=& 
\int {d\theta_p \over 2\pi} {d\theta_h \over 2\pi}
\lb j_L (x,t) | \theta_h \theta_p \rb \lb \theta_p \theta_h 
| j_L (0)\rb \cr\cr
-{K \over (ix+v_F\tau)^2} &=& {c^2\mu^2\over 4\pi^2} 
\int d\theta_h d\theta_p e^{-\mu (ix+v_F\tau) (e^{\theta_h}+e^{\theta_p})}
e^{\theta_h+\theta_p} \cr \cr
&=& {c^2 \over 4\pi^2 (ix+v_F\tau)^2}.
\end{eqnarray}
On the other hand, $c$ is related to $S_R$, as integrating the above
matrix element gives the value of $S_R$ on the one-particle
states:
\begin{eqnarray*}
2\pi S_R \delta (\theta_1-\theta_2) &=& 
\int dx \lb \theta_1 | M^3_0 (x)| \theta_2 \rb = 
\int dx \lb \theta_1 |{j_L(x) \over 2\pi} |\theta_2 \rb \cr\cr
&=& -i \int {dx\over 2\pi}
\mu c e^{(\theta_2+\theta_1)/2}e^{i\mu x (e^{\theta_2}-e^{\theta_1})}\cr\cr
&=& -i c \delta (\theta_1-\theta_2).
\end{eqnarray*}
Thus $c= i2\pi S_R$ and so 
\begin{equation}\label{exxxiv}
K = S_R^2.
\end{equation}
Given $S_R$ is described by the integral equation (\ref{exxxi}),
$K$ follows immediately. The results are plotted in Figure 3.

It is straightforward to find explicit power-series expansions for
$K$ and $v_F$ valid for $H$ near $\Delta$.
We begin by computing $\ep (\theta )$.  For $H$ slightly larger than $\Delta$,
we can expand $\ep(\theta )$ around $\theta = 0$:
\begin{equation}\label{exxxv}
\ep (\theta ) = d_0 + d_2 \theta^2 + {\cal O}(\theta^4) .
\end{equation}
Plugging this into (\ref{exxiv})
demands $d_0$ and $d_2$ satisfy
\begin{eqnarray}\label{exxxvi}
d_0 &=& (\Delta - H)(1+ 2\Gamma_{++} (0) \theta_F ) + 
\Delta {\Gamma_{++} (0)\over 3}\theta_F^3 + {\cal O}(\theta_F^4);\cr\cr
d_2 &=& {\Delta \over 2} - 2\theta_F \Gamma^2_{++} (0) (\Delta - H) + 
{\cal O}(\theta_F^4),
\end{eqnarray}
where $\Gamma_{++} (0) = 1/\pi^2$.
To determine $\theta_F$ we apply the condition
$ \ep (\theta_F) = 0$,
resulting in 
\begin{equation}\label{exxxvii}
\theta_F = \sqrt{2({H\over \Delta}-1)} + 
{4\over 3}\Gamma_{++} (0) ({H\over \Delta}-1)
+ {\cal O}(({H\over \Delta}-1)^{3/2}).
\end{equation}
Similarly we can show $\rho (\theta)$ to be
\begin{eqnarray}\label{exxxviii}
\rho (\theta ) &=& \rho_0 + \rho_2 \theta^2 ,
\end{eqnarray}
where
\begin{eqnarray}\label{exxxix}
2\pi\rho_0 &=& \Delta (1+2\Gamma_{++} (0)\theta_F 
+ 4\theta_F^2\Gamma^2_{++}(0)) 
+ {\cal O}(\theta_F^3);\cr\cr
2\pi\rho_2 &=& 
{\Delta\over 2} - 2\theta_F\Gamma^2_{++}(0)\Delta + {\cal O}(\theta_F^2).
\end{eqnarray}
We can find the renormalized spin similarly.

These expansions give the
power-series expansions for $K$ and $v_F$ to be
\begin{eqnarray}\label{exl}
K &=& 1 + {2^{5/2}\over \pi^2}\left({H\over\Delta}-1\right)^{1/2} + 
{88\over 3\pi^4}\left({H\over\Delta}-1\right) + {\cal O}\left({H\over\Delta}-1\right)^{3/2};\cr\cr
v_F &=& v_{F_0}
\left[\sqrt{2}\left({H\over\Delta} - 1\right)^{1/2} - {8\over 3\pi^2}\left({H\over\Delta}-1\right)
+{\cal O}\left({H\over\Delta}-1\right)^{3/2}\right].
\end{eqnarray}
We have restored the bare spin-wave velocity
$v_{F_0}$ (earlier set to one).

We see that as a function of the magnetic field, the Luttinger
parameter $K$ is 1 at threshold $H=\Delta$, and
then increases with increasing field strength.  That
$K=1$ at $H=\Delta$ is a universal result already
obtained by \cite{affleck1a,haldane2}.  It is independent
of the exact nature of the excitations that fill the
ground state at finite magnetization.  In the low
magnetization regime, the excitations of the ground
state interact only weakly due to their low density.
Thus to a zeroth order approximation, they can be treated
as free fermions with a corresponding Luttinger parameter
of $K=1$.
Thus in the analysis of antiferromagnetic spin
ladders \cite{gia}, $K=1$ at $H=\Delta$ is again found even though
$K$ decreases (opposite to the behaviour of the spin-1 chains)
with subsequent increases in $H$.
Similarly in mapping the spin chain onto an interacting Bose gas 
\cite{Takahashi},
one arrives at $K=1$ for $H\sim \Delta$.  Here however
$K$ at least increases with increasing field.  By fine tuning
the strength of the interactions in the gas, one could in principle
come close to reproducing the behaviour as predicted by
the O(3) NLSM.

We illustrate this in more detail.  The equations governing
the energy and renormalized spin of the Bose
gas are \cite{Korepin} (in the case of the Bose
gas the quantity equivalent to $S_R$ is the renormalized charge)
\begin{eqnarray}\label{exli}
\ep (\theta ) &=& \theta^2 - H + \int^{\theta_F}_{-\theta_F} 
d\theta' \ep(\theta') \Gamma_{++}(\theta - \theta');\cr\cr
S_R &=& 1 + \int^{\theta_F}_{-\theta_F} S_R(\theta ') \Gamma_{++} 
(\theta_F-\theta');\cr\cr
\Gamma_{++}(\theta ) &=& {c \over \pi} {1 \over c^2 + \theta^2}.
\end{eqnarray}
Here $c$ is the strength of the interactions in the Bose gas.
If $c=\pi$, we see the equations of the Bose gas are nearly identical
to those of the $O(3)$ NLSM.  The sole difference (apart from
some trivial shifts and rescalings) lies in the dependence
of the bare energy upon $\theta$, i.e. 
$\ep_0^{\rm Bose} (\theta ) = \theta^2 - H$ while
$\ep_0^{\rm NLSM} (\theta ) = \Delta \cosh (\theta) - H$.
At low energies the two are identical, but diverge at
higher scales.

\section{Scaling Behavior at $H > \Delta$}

Using the results for $K$ and $v_F$ derived in the
last section, we describe the scaling behavior present
in the magnetized phase of the spin chain.

\subsection{Susceptibility and Spin Conductance}

We first consider the computation of the magnetic susceptibility.
To take into account the effects of a magnetic field we
add a term to the Lagrangian of the form
\begin{eqnarray}\label{exlii}
\delta{\cal L} &=& H \int dx dt M_z\cr\cr
&=& H \int dx dt {K\over \pi v_F}\partial_t \Phi .
\end{eqnarray}
Here we have used that the oscillating term of $M_z$ vanishes with
the integration over $x$.  
Doing so the relevant correlator in computing the susceptibility
is then
$$
\lb M_z \partial_t \Phi \rb.
$$
This 
correlator can be evaluated in the bosonic formulation
of the  Luttinger liquid giving
\begin{equation}\label{exliii}
\langle M_z (x,\tau ) {K\over \pi v_F} \partial_t \Phi (0)\rangle 
= {K T^2 \over 4 v_F^2}
\bigg[ {1\over \sinh^2 ({i\pi T\over v_F}z)}
+ {1\over \sinh^2 ({i\pi T\over v_F}\bar{z})}\bigg] .
\end{equation}
Here we have evaluated it at finite temperature.
By taking the appropriate analytical continuations,
we can arrive at the retarded correlator necessary to compute
the susceptibility, $\chi (H)$.  We so find
\begin{equation}\label{exliv}
\chi(H>H_c) = 
\langle M_z {K\over \pi v_F}\partial_t \Phi 
\rangle_{\rm retarded}|_{\omega=0,k=0}
=  {K(H) \over \pi v_F (H).}
\end{equation}
We point out that the static susceptibility is independent of temperature.

As we are able to compute the susceptibility of the spin
chain directly from the thermodynamics of the $O(3)$ NLSM,
we are able to perform
a non-trivial check on the correctness of this 
calculation.
The susceptibility, $\chi$, at T=0
is given by taking two derivatives of the energy,
as displayed in (\ref{exxv}) and (\ref{exxvi}).
Expanding this in a power series as explained at the end of the
last section gives the energy per unit length to be
\begin{equation}\label{exlv}
E(H) = -{\Delta^2\over 2\pi}
\left[ {2^{5/2}\over 3}\left({H\over\Delta} -1\right)^{3/2}
+ {16\over 3\pi^2}\left({H\over\Delta} -1\right)^2 
+{\cal O}\left({H\over\Delta}-1\right)^{5/2}\right],
\end{equation}
so that the susceptibility is
\begin{equation}\label{exlvi}
\chi (H) = -\partial_H^2 E(H) = 
{\sqrt{2}\over 2\pi} \left({H\over\Delta} -1\right)^{-1/2}
+ {16\over 3\pi^3} +{\cal O}\left({H\over\Delta}-1\right)^{1/2}
= {K(H) \over \pi v_F(H)}.
\end{equation}
We see that we recover the susceptibility
as derived from the above Kubo formula.  

We can also compute the spin conductivity.  The spin conductivity
measures the resulting spin current arising from a spatially varying
magnetic field and is defined via
$$
j_s = \sigma_s \nabla H.
$$
The spin current operator has the general form
$$
j_s = - v_F{K\over\pi}\partial_x \Phi + {\rm term~dependent~upon~}M.
$$
However the latter term in $j_s$ involving $M$ will not couple
to $\delta {\cal L}$ in (\ref{exlii}).  As such the spin
conductance is given simply from a Kubo formula,
\begin{eqnarray}\label{exlvii}
{\rm Re} \sigma_s (\omega, H) &=& \lim_{k\rightarrow 0} {1\over k}
{\rm Im} \lb j_s
\bigg(-{K\over v_F \pi}\partial_t\Phi\bigg) \rb_{\rm ret} (\omega , k)\cr\cr
&=& \lim_{k\rightarrow 0} {K^2 \over \pi^2}{1\over k}
{\rm Im} \lb\partial_x\Phi
\partial_t\Phi \rb_{\rm ret} (\omega , k)\cr\cr
&=& v_F K(H) \delta(\omega ).
\end{eqnarray}
This result is valid at any temperature within the Luttinger liquid
framework.

\subsection{Inelastic Neutron Scattering}

We can make a series of predictions relevant for inelastic
neutron scattering experiments with $H > H_c$.  For scattering
near wave-vector $k=\pi$, the experiments probe the single particle
spectral weight of the field, $n^a(x,t)$.  
The scattering cross section is given in terms of  
the correlation function,
\begin{equation}\label{exlviii}
\sigma (\omega , k) \propto \int d \omega d k\, e^{-i\omega t + ikx} 
\sum_a \lb n^a (x,t) n^a(0,0)\rb.
\end{equation}
Because $n_3$ is massive, its spectral weight is exponentially
suppressed at low energies
and temperatures
and for such cases we need only consider the contribution of
the correlator $\lb n^+ n^-\rb$:
\begin{equation}\label{exlix}
\sigma (\omega , k) \propto \int dx dt\, e^{-i\omega t + ikx} 
\lb n^+ (x,t) n^-(0,0)\rb.
\end{equation}
We first examine the $T=0$ behavior of this correlator.

\begin{figure}
\centerline{\rotatebox{-90}{\epsfxsize=3.35in
\epsfysize=3.5in\epsffile{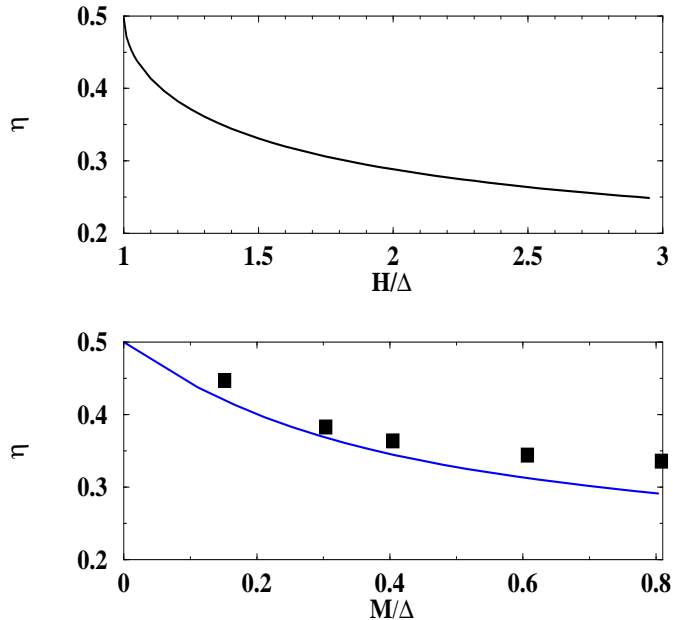}}}
\caption{Plots of the scaling exponent $\eta$ for the transverse 
spin-spin correlator, $\langle S_x (x) S_x (0)\rangle \sim |x|^{-\eta}$.  
In the top panel $\eta$ is plotted versus the applied magnetic field
while in the bottom panel the $\eta$ is plotted versus the magnetization
of the ground state.  In this latter graph we also plot the values
(squares) of $\eta$ found in Campos Venuti et al. through 
a density matrix renormalization group analysis.}
\end{figure}

From the relation (\ref{exix}), the 
correlator $\lb n^+ n^- \rb$ equals
\begin{eqnarray}\label{el}
\lb n^+ (x,\tau ) n^- (0,0) \rb &=& {1\over (x^2 + v^2_F \tau^2)^{1/4K}}
+ 2c_1^2 \cos (2\pi M x) {1\over (x^2 + v^2_F \tau^2)^{K+1/4K}}.
\end{eqnarray}
The exponent, $\eta = 1/2K$, governing the leading
piece of the transverse spin-spin correlator
is plotted in the top panel of Figure 4 as a function of magnetic
field.  We see that its $H = \Delta^+$ value is $\eta = 1/2$, equal
to that of free fermions.  For values of $H$ slightly larger than
$\Delta$, $\eta = 1/2K$ takes the form
\begin{equation}\label{eli}
\eta = {1\over 2K} = {1\over 2}\bigg(1-{2^{5/2}\over\pi^2}
( {H\over\Delta} -1)^{1/2}
+ {8\over {3\pi^4}}({H\over\Delta} -1) + 
{\cal O}({H\over\Delta} -1)^{3/2}\bigg).
\end{equation}
In the bottom panel of Figure 4 we plot $\eta$ versus the magnetization
of the ground state.  We also compare our computation to that of
numerical simulations of a lattice 
integer-spin chain, $H = J\sum_i S_i\cdot S_{i+1}$ 
done by Campos Venuti et al.\ \cite{numerics}.  The agreement
between the two is reasonable but not overwhelming.  At larger values of
the magnetization, $M$, this is to be expected.  For a lattice spin
chain the magnetization must saturate at some critical value
of the applied magnetic field, $H_c \sim J$.  In contrast the
magnetization of the $O(3)$ sigma model has no such upper bound.  Thus we
must work at values of the applied field where the magnetization
of the system is small or alternatively, $H \ll J$.  Given
that $\Delta \sim .4J$ in a spin 1 chain, this limit can only be ambiguously
met at best.
At smaller values of the magnetization, the disagreement may arise
from differences between the lattice theory and its continuum version.
Nonetheless, we find reasonable agreement.

At finite temperature we can easily determine the form of $\lb n^+ n^-\rb$.
Via a conformal transformation we have,
\begin{eqnarray}\label{elii}
\lb n^+ (x,\tau ) n^- (0,0) \rb &=&
\bigg({\pi^2\over v_F^2\beta^2}\bigg)^{1/4K}
{1\over |\sinh T\pi (x/v_F + i\tau )|^{1/2K}}\cr\cr
&& \hskip -.5in 
+ 2c_1^2\cos (2\pi M x) \bigg({\pi^2\over v_F^2\beta^2}\bigg)^{K+1/4K}
{1\over |\sinh T\pi (x/v_F + i\tau )|^{2K+1/2K}}.
\end{eqnarray}
Analytically continuing and then Fourier transforming thus
 gives us an expression for the cross section at finite T,
\begin{eqnarray}\label{eliii}
\sigma (\omega , k) &=& {1\over 2v_F} 
\bigg({\pi \over \beta v_F}\bigg)^{-2+1/2K} 
f_{1/4K}({\beta\over\pi}\omega,{\beta v_F\over\pi}k)
+ c_1^2 {1\over 2v_F}\bigg({\pi \over \beta v_F}\bigg)^{-2+K+1/4K}  \cr\cr
&& \hskip -.75in 
\times \bigg(f_{K+1/4K}({\beta\over \pi}\omega ,{\beta v_F\over\pi}(k+2\pi M))
+ f_{K+1/4K}({\beta\over \pi}\omega,{\beta v_F \over \pi}(k-2\pi M))\bigg),
\end{eqnarray}
with $f_\gamma(x,y)$ equal to \cite{orgad},
\begin{eqnarray}\label{eliv}
f_\gamma(x,y) &=& h_\gamma({1\over2}(y-x))h_\gamma({1\over2}(y+x));\cr\cr
h_\gamma (x) &=& 
{\rm Re}\big[ (2i)^\gamma B({\gamma-ix \over 2},1-\gamma)\big],
\end{eqnarray}
where $B$ is the beta function, $B(x,y) = \Gamma(x)\Gamma(y)/\Gamma(x+y)$.
We so obtain a scaling form for $\sigma (\omega , k)$ with $f(x,y)$ a
universal function.  The scaling form two parts: one relevant for wavevectors
$k$ near $0$ and one yielding a contribution for $k$ near the
incommensurate wave vector $2\pi M$ (understanding that all
wavevectors involving $n^\pm$ have been shifted by $\pi$).

We can analyze the above expression for 
$\sigma (\omega ,k)$ in the small and large $T$ limits.
In the case $T\ll M$ and $\omega , k \ll M$, 
we can safely ignore the contribution
of the second term in (\ref{eliii}).   We then find
\begin{eqnarray}\label{elv}
\sigma (\omega ,k) &\sim& {2\pi^2\over v_F \Gamma^2(1/4K)}
\Theta(-\omega-v_Fk)\Theta(v_Fk-\omega)
\bigg({\omega^2/v_F^2-k^2 \over 16}\bigg)^{-1+1/4K}
; ~~~ T \ll \omega,k;\cr\cr
\sigma (\omega ,k) &=& {2\over v_F}
\bigg({2\pi\over\beta v_F}\bigg)^{-2+1/2K}\cos^2({\pi\over 8K})
B^2({1\over 8K},1-{1\over 4K}); ~~~ T \gg \omega,k ;\cr\cr
&\sim& T^{-2+1/2K}.
\end{eqnarray}
In the first case, $T \ll \omega, k$, the leading term
is temperature independent.
For the behavior in the second case to be observed, we need $\Delta \gg T$.
Otherwise thermal excitations involving the other two bands, $(S_z = 0,-1)$,
would alter the Luttinger liquid behavior of the ground state.
For wavevectors near $k\pm 2\pi M$ the second term
in (\ref{eliii}) dominates and we find instead
\begin{eqnarray}\label{elva}
\sigma (\omega ,k\pm 2\pi M) &\sim& c_1^2{2\pi^2\over v_F \Gamma^2(1/4K+K)}
\Theta(-\omega-v_Fk)\Theta(v_Fk+\omega)\cr\cr
&& \hskip 1.75in \times \bigg({\omega^2/v_F^2-k^2 \over 16}\bigg)^{-1+K+1/4K}
; ~~~ T \ll \omega,k;\cr\cr
\sigma (\omega ,k\pm 2\pi M) &=& c_1^2 {2\over v_F} 
\bigg({2\pi\over\beta v_F}\bigg)^{-2+2K+1/2K}
\cos^2(\pi({1\over 8K} + {K\over2}))\cr\cr
&& \hskip 1.75in \times
B^2({1\over 8K}+{K \over 2},1-K-{1\over 4K}); ~~~ T \gg \omega,k ;\cr\cr
&\sim& T^{-2+2K+1/2K}.
\end{eqnarray}
We thus see the power law dependencies of $\sigma (\omega ,k)$ near 
$k \sim 2\pi M$ differ from those near $k \sim 0$.

For scattering near the wave-vector, $k=0$, the spectral
weight of the magnetization operator is probed.  It is this operator
that creates/destroys excitations near $k=0$.  At low energies
and temperatures the cross section, $\sigma$, is given by
\begin{equation}\label{elvi}
\sigma (\omega , k) \propto \int dt dx e^{-i\omega t + ikx} 
\lb M_z (x,t) M_z(0,0)\rb.
\end{equation}
The other two spin components of the magnetization are massive and
do not contribute at low energies.  By the fluctuation-dissipation
theorem, we can recast $\sigma$ in terms of the imaginary
piece of the corresponding retarded correlator:
\begin{equation}\label{elvii}
\sigma (\omega , k) \propto - {2\over 1 - e^{-\beta \omega}} 
{\rm Im} \lb M_z M_z\rb_{\rm ret.}(\omega , k) .
\end{equation}
In the case that $\omega , k \ll M$, we obtain
\begin{equation}\label{elviia}
\sigma (\omega , k) \propto  {K \over v_F} {\omega \over 1 - e^{-\beta \omega}} 
(\delta (\omega + v_F k)+\delta(\omega- v_F k)).
\end{equation}
For wavevectors near $2\pi M$ we find instead
\begin{eqnarray}\label{elviib}
\sigma (\omega , k\pm 2\pi M) &\sim & -{c^2 \over 1 - e^{-\beta\omega}}
\Gamma^2(1-K)\sin(\pi (K-1)){\sin(\pi K)\over v_F}
\bigg({\omega^2-k^2 \over 4}\bigg)^{K-1}
\cr\cr
&& \hskip 1in \times ({\rm sgn}(w+k)+{\rm sgn}(w-k)); 
~~~ T \ll \omega , k;\cr\cr
\sigma (\omega , k\pm 2\pi M) &\sim & -{2c^2\omega \over 1 - e^{-\beta\omega}}
\bigg({2\pi\over \beta v_F}\bigg)^{2K-3} {\sin (\pi K)\over v_F}
B^2({K\over 2},1-K)\psi (K/2); ~~ T \gg \omega , k ;
\end{eqnarray}
where $\psi (x) = \partial_x \log (\Gamma (x))$.

\subsection{NMR Relaxation Rates}

As with the inelastic neutron scattering cross section,
we can make a series of predictions for NMR relaxation rates.
The NMR relaxation rate is given in
terms of the spin-spin
correlation function \cite{sagi}:
\begin{equation}\label{elviii}
{1\over T_1} = \sum_{\alpha = 1,2 \atop \beta = 1,2,3}
\int {dk\over 2\pi} A_{\alpha\beta}(k)A_{\alpha\gamma}(-k)
\lb n^\beta n^\gamma \rb (k,\om_N),
\end{equation}
where $\om_N = \gamma_N H$ is the nuclear Lamour frequency with 
$\gamma_N$ the nuclear gyromagnetic ratio and the $A_{\alpha\beta}$
are the hyperfine coupling constants.  In the above we have assumed $H$ is
aligned in the 3(z)-direction.
We assume that the hyperfine couplings are independent of the wavevector
$k$.  Hence
\begin{equation}\label{elix}
{1\over T_1} \propto A_{\alpha\beta}A_{\alpha\gamma}
\lb n^\beta n^\gamma \rb (x=0,\om_N \sim 0).
\end{equation}
We now consider the correlator $\lb n n \rb$ in more detail.
Specifically we will compute the contributions to $T_1^{-1}$
coming from the different branches of low energy
excitations.

If the hyperfine couplings are such that the 
transverse fields are dominant we
obtain for $1/T_1$,
\begin{eqnarray}\label{elx}
{1\over T_1} &\sim& \lim_{\omega \rightarrow 0}\int dt e^{i\omega t}
\lb n_+ (t,x=0)n_-(0) \rb \cr\cr
&=& \lim_{\omega \rightarrow 0}
{1\over 2v_F}\bigg({\pi \over \beta v_F}\bigg)^{-1+1/2K}
h_{1/2K}(\beta\omega/\pi ) + 
{c_1^2\over v_F}\bigg({\pi \over \beta v_F}\bigg)^{-1+2K+1/2K}
h_{1/2K+2K}(\beta\omega/\pi ),
\end{eqnarray}
where $h_{1/2K}$ is as defined in (\ref{eliv}).
In the low temperature limit the second term is subdominant
and we have 
\begin{equation}\label{elxi}
{1\over T_1} \propto {1\over 2v_F}\bigg({\pi \over \beta v_F}\bigg)^{-1+1/2K}
h_{1/2K}(0) .
\end{equation}
In the low density limit (i.e. $H \sim \Delta^+$), 
we have $1/T_1 \sim T^{-1/2}$.
In this limit $1/T_1$ has a dependence upon $T$
identical to that of antiferromagnetic ladders \cite{gia,chitra}.
However as $H$ is increased, the power law dependencies 
of the two cases diverge.

If on the other hand the hyperfine couplings
are such that the contribution from $\lb M_z M_z \rb$ is important
we find,
\begin{eqnarray}\label{elxii}
{1\over T_1} &\sim& \lim_{\omega \rightarrow 0}\int dt e^{i\omega t}
\lb M_z (t,x=0)M_z(0) \rb,  \cr\cr
&=& {2 K T \over v_F^2} +  c^2 \bigg({\pi\over \beta v_F}\bigg)^{2K-1}
h_{2K} (0).
\end{eqnarray}
As $K>1$ for any finite H in excess of $\Delta$,
the first of the two terms dominates.

\subsection{Validity of Luttinger Liquid Picture}

\begin{figure}
\hskip.7in \rotatebox{-90}{\epsfxsize=3.5in\epsffile{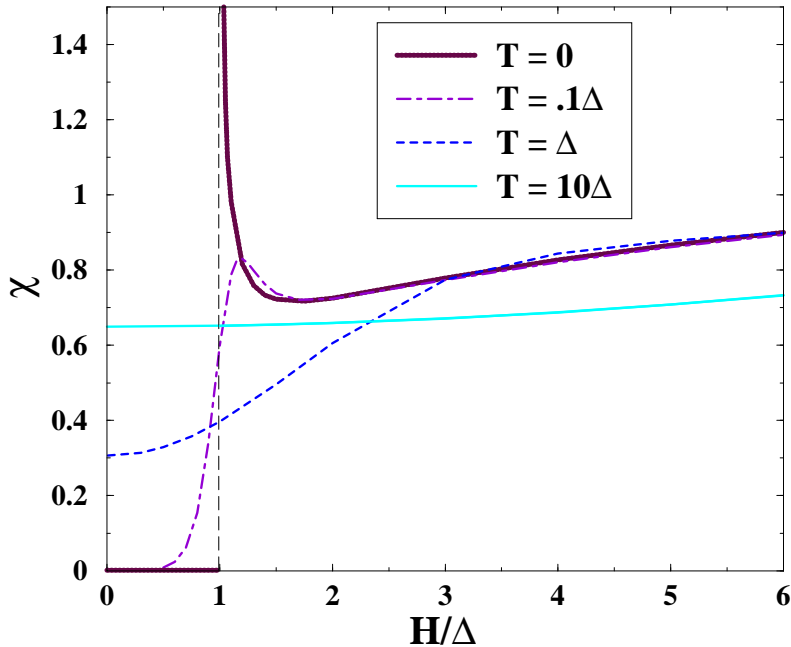}}
\caption{Plots of the susceptibility as a function of applied field
for a variety of temperatures.}
\end{figure}

The Luttinger liquid picture we have developed is precise only at 
zero temperature.  Nonetheless we have argued that this behavior
will persist to some degree at finite temperature.  We are in position
to analyze qualitatively at least whether this is indeed true.
To this end, we compute the susceptibility at finite temperature
and finite magnetic field.  We do so using a more sophisticated
formalism than presented previously: the
thermodynamic Bethe ansatz.  For the $O(3)$ sigma model,
the appropriate equations were originally given in \cite{tsvelik}
but can also be found in \cite{rmk,meTBA}.\footnote{Apropos of
nothing, we have also computed the susceptibility
at large temperatures but zero field.
In the zero-field large-$T$ limit,
Sachdev and Damle \cite{damle2} provide a high temperature
computation of the
susceptibility predicated on integrating out higher
Matsubara frequencies.  To subleading order it is given by
\begin{equation}\nonumber
\chi(T) = {1\over 3\pi} (\log ({32\pi e^{-2-\gamma} T \over \Delta})
+ \log(\log({8 T \over e\Delta})) + {\cal O}({\log(\log(T))\over\log(T)}).
\end{equation}
From a numerical analysis of the TBA equations at
temperatures in the range, $T \sim 10^3\Delta-10^7\Delta$, we find agreement
with this analytical form at the $1\%$-level.  To obtain 
this agreement it was important to include the subleading term.}

Plots of the exact susceptibility for various temperatures derived
from these equations are given in Figure 5.   
Note in particular how the peak at $H=\Delta$ appears
as the temperature is lowered.  At $T=0$, we see from (\ref{exlvi}) that
this turns into a square-root singularity.  The divergence of
the susceptibility from its $T=0$ value is a reasonable indication
of the persistence of the Luttinger-liquid picture at finite temperature
as $\chi$ can be given directly in terms of $K$ and $v_F$ 
(see (\ref{exxiv})).  Thus we expect finite
temperatures to be destructive of Luttinger liquids associated with
fields, $H$, only slightly in excess of the gap where a finite temperature
drastically rounds off the square root singularity appearing in $\chi$.
However at higher fields where the density of magnons in the ground
state is larger and so presumably more robust against small temperature
perturbations, the susceptibility is equal to its zero temperature value.
In this case we expect the $T=0$ Luttinger liquid picture to remain
valid.

We note as an aside the square root singularity in $\chi$ appearing
in Figure 5 at $T=0$ is a generic feature.  It appears in
any model (not only spin chains) with a gap and a low-energy quadratic
dispersion relation.  Its appearance is not related in any way to
the $O(3)$ NLSM being integrable (although the integrability allows
us to compute exactly the coefficient of the divergence).
We thus expect this square-root divergence to appear also in spin
chains with large easy-axis anisotropies.   Indeed the rounded
finite temperature
counterpart of this divergence has already been seen in NENP
\cite{ajiro}.

\begin{figure}
\centerline{\rotatebox{-90}{\epsfxsize=3.0in\epsffile{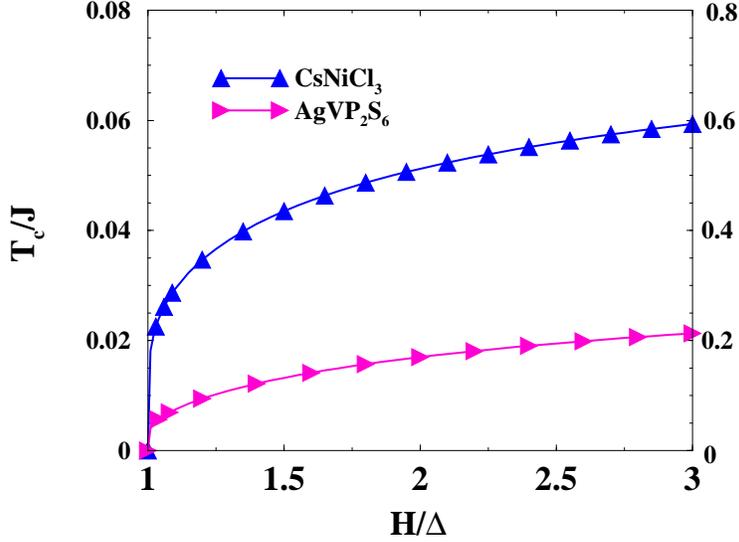}}}
\caption{A plot of the dependence of the 3D ordering
temperature, $T_c/J$, as a function of applied field, $H$,
for two different spin chain compounds.  The relevant
values of $J$ for two materials are
$J_{CsNiCl_3} \sim 16K$ and 
$J_{AgVP_2S_6}\sim 320K$.  The corresponding values of inter to
intrachain anisotropies are $J'/J \sim .017$ ($CsNiCl_3$)
and $J'/J \sim 10^{-5}$ ($AgVP_2S_6$).  The left ordinate scale
corresponds to $AgVP_2S_6$ while the right scale corresponds
to $CsNiCl_3$.}
\end{figure}

Beyond the effects of finite temperature, the Luttinger liquid
phase can be destroyed by the presence of intrachain couplings.
In mean field theory, three dimensional order due to such couplings
develops at a transition
temperature, $T_c$, given by
\begin{equation}\label{elxiii}
1 = |a J' \chi_\perp (q=0,\omega = 0, T=T_c)|,
\end{equation}
where $\chi_\perp$ is the staggered susceptibility of a single chain,
\begin{equation}\label{elxiv}
\chi_\perp = \int dx dt e^{i\omega t - ikx} 
\langle n^+(x,t)n^-(0)\rangle_{\rm retarded},
\end{equation}
and $J'$ is the intrachain coupling.  Here we explicitly
include the lattice cutoff, $a^{-1} \sim J$.  $q$ is the deviation
from the wavevector, $\pi$.
From the expressions for $n^\pm$ given in (\ref{exix}), 
we obtain, similar to \cite{schultz},
\begin{eqnarray}\label{elxv}
\chi_\perp (q,w) &=& - ({2\pi a \over \beta v_F})^{(-2+1/(2K))} 
{\sin ({\pi\over 4K}) \over v_F}
\bigg[B({1\over 8K} - i{v_F\omega_-\over 4\pi T},1 - {1\over 4K})
B({1\over 8K} - i{v_F\omega_+\over 4\pi T},1 - {1\over 4K})\bigg]\cr\cr
&&\hskip -.75in - c_1^2 ({2\pi a\over \beta v_F})^{(-2+2K+{1\over 2K}))} 
{\sin ({\pi (K+{1\over 4K})}) \over v_F}\bigg[
B({K\over 2}+{1\over 8K} - i{v_F(\omega_--2\pi M)\over 4\pi T}
,1 - K-{1\over 4K})\cr\cr
&& \hskip .70in \times
B({K\over 2}+{1\over 8K} - i{v_F(\omega_++2\pi M)\over 4\pi T},1-K-{1\over4K})
+ (M \rightarrow -M)\bigg] ,
\end{eqnarray}
where $\omega_\pm = (\omega/v_F \pm k)$.  
Combining this expression with the mean condition leads to a transition
temperature given by \cite{imry}
\begin{eqnarray}\label{elxvi}
{T_c \over J} &=& {1\over \pi} {v_F \over v_{F0}}
({1\over 2}{J' \over J }{v_{F0}\over v_F}\delta)^{2K/(4K-1)};\cr\cr
\delta &\equiv &  \sin({\pi\over 4K})
B^2({1\over 8K},1 - {1\over 4K}).
\end{eqnarray}
In deriving this expression we have ignored the 
term in $\chi_\perp$ appearing in (\ref{elxvi}) dependent upon $M$ whose 
contribution is negligible for small $T$.

Plotted in Figure 6 is the dependence of $T_c$ upon $H$
for the spin chain compounds $CsNiCl_3$ and $AgVP_2S_6$.  
We see that as $H$ is increased there is a corresponding
increase in $T_c$.  Thus at a fixed temperature and
sufficiently large interchain coupling, $J'$, we expect
that as $H$ is increased there will be a transition to a
state with long range order.  We thus see that 
the Luttinger liquid phase is stable at a given temperature
for smaller fields.

As discussed in
the introduction, field induced long range order has
been observed in the spin chain material, NDMAP \cite{honda}.
Although this compound has a large single ion anisotropy,
we make an attempt to compare experimental observations
with our analysis.  From inelastic neutron scattering experiments,
the spin coupling, $J$, of NDMAP is $J=26.5K$ \cite{Zheludev}.
For a field applied along the chain
(perpendicular to the easy plane defined by the anisotropy),
the corresponding gap $\Delta$ is given in terms of $J$
and $D=.3J$, the strength of the single ion anisotropy, 
by \cite{compgap},
$$
\Delta \sim .4J - .66D = 5.3K = 3.75T,
$$
with the Land\'e g factor equal to $g=2.1$. 
In order to obtain a reasonable fit with the experimental
observations we take $J'/J = 2\times10^{-4}$.  
The results are plotted in Figure 7.
The value used for $J'/J$ is considerably smaller (by a factor of 6)
than the value employed in \cite{honda}.  These authors, however, were
able to match the experimental data by assuming $\chi_\perp$ took the
form $\chi_\perp \sim AT^{-2+1/2K}$ with $A$ some field independent constant.
We however find that $A$ is both field dependent and generally
larger in magnitude than the value of $A$ used in \cite{honda}.
To compensate in matching the data, we need to take $J'/J$ to
be smaller in magnitude.

\begin{figure}
\centerline{\rotatebox{-90}{\epsfxsize=3.0in\epsffile{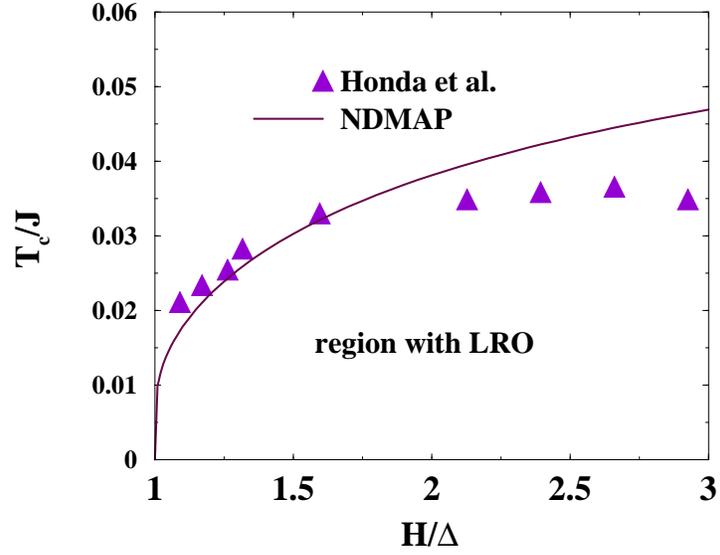}}}
\caption{A plot of the dependence of the 3D ordering
temperature, $T_c/J$, as a function of applied field, $H$,
for the spin chain compound, NDMAP.}
\end{figure}

At fields not far in excess of the gap, we find reasonable agreement
between the data and the theoretical computation.  At larger
values of the field, $H$, the computed $T_c$ exceeded the observed value.
This is not surprising.  As we saw in our computation of the exponent
$\eta$, the assumptions underlying our computation fail
to accurately describe the physics at large fields.  With $H/\Delta > 2$,
we enter a region where the magnetization per site of the system
is some significant fraction of its maximal value of 1, a regime
our method, which puts no bound upon the magnetization, is unable to
handle properly.

The destruction of the Luttinger liquid phase due to an applied
magnetic field occurs not only for spin-1 chains but for antiferromagnetic
spin-1/2 ladders \cite{gia}.  
Unlike the result found in the case of two antiferromagnetically
coupled spin-1/2 chains \cite{gia}, we find however $T_c$ is strictly
monotonic as a function of applied field.  The difference in
the two cases is again that we have $K>1$ and increasing with
field while
in the case of antiferromagnetic spin ladders, $K<1$ and decreasing
with field.
There is an additional instability that affects antiferromagnetic spin
ladders, an instability to the formation of a spin-Peierls state
due to coupling to the lattice \cite{nagaosa}.  
However spin-1 chains are thought to be
robust against this perturbation \cite{nagaosa}.

\vfill\eject

\appendix

\section{Analysis of the Ground State at T=0}

In this section we determine the characteristics of the zero
temperature ground state in a magnetic field, $H$, exceeding
the gap.
At the heart of this analysis lies the insistence
that the ground 
state wave function be compatible with the scattering amplitude, $S_{++}$,
as defined in (\ref{exxi}).
This amplitude characterizes the
wave function, $\psi (\theta_1,\ldots,\theta_N )$,
of a ground state with $N$ particles through the application of periodic
boundary conditions.  In allowing the i-th particle, $\theta_i$, to traverse
the entire length, $L$, of the system and so commute with the other $N-1$
particles, the wave function picks up a phase
\begin{eqnarray}\label{eAi}
\psi (\theta_1,\ldots,\theta_N ) 
&\rightarrow& \phi (\theta_i )\times \psi (\theta_1,\ldots,\theta_N ) ;\cr\cr
\phi (\theta_i ) &=& 
\prod^N_{j\neq i} S_{++} (\theta_i-\theta_j)e^{iL\Delta\sinh(\theta_i)}.
\end{eqnarray}
The first term is a product of scattering amplitudes arising from the 
i-th particle scattering with the remaining $N-1$ particles.  The second
term comes from the bare momentum carried by the i-th particle.  With
periodic boundary conditions, we must have
$$
\phi (\theta_i ) = 1 .
$$
Taking logarithms of this constraint leads to the quantization condition,
\begin{equation}\label{eAii}
N_i = {L \Delta \over 2\pi}\sinh (\theta_i ) + \sum_{j\neq i} {1\over 2\pi i}
\log S_{++} (\theta_{ij}),
\end{equation}
where $N_i$ is the quantum number characterizing the rapidity, $\theta_i$,
(i.e. $\log (1) = i2\pi N_i$).
As we approach the continuum limit, we can then write a difference
equation between successive integers, $N_i$ and $N_{i+1} = N_i +1$,
\begin{equation}\label{eAiii}
1 = {L\Delta \over 2\pi} \cosh(\theta_i) \delta\theta_i + \sum_{j\neq i}
{1\over 2\pi i}(\partial_\theta \log S_{++}) (\theta_{ij})\delta\theta_i; ~~ 
\delta\theta_i = \theta_{i+1}-\theta_i.
\end{equation}
The density of states per unit length at $\theta_i$ is
defined as $n(\theta_i) \equiv 1/L\delta\theta_i$.  
With this we can rewrite the above
in the continuum limit as,
\begin{eqnarray}\label{eAiv}
\rho (\theta ) + \tilde\rho (\theta ) &=& {\Delta \over 2\pi} \cosh (\theta )
+ \int d \theta' \rho (\theta ') \Gamma_{++} (\theta-\theta'); 
\end{eqnarray}
where $\rho$ is the density of occupied states
and $\tilde\rho$ is the density of unoccupied states, so that
$\rho+\tilde\rho = n$.  

At zero temperature, the equation for $\rho$ can be simplified.
With $T=0$, the ground state has a Fermi surface at
$\theta_F$.  
Excitations with rapidities, $|\theta| > \theta_F$,
do not appear in the ground state.  Thus
\begin{eqnarray}\label{eAv}
\rho (\theta ) &=& \Bigg\{ {n (\theta ), ~~~ |\theta| < \theta_F;
\atop 0, ~~~~~~  |\theta| > \theta_F ;}\cr\cr
\tilde\rho (\theta ) &=& \Bigg\{ {0, ~~~ |\theta| < \theta_F;
\atop n (\theta ), ~~~~  |\theta| > \theta_F ;}
\end{eqnarray}
and we obtain equation (\ref{exxii}) for $\rho$ at zero temperature.

We have computed the interacting density of states.  We can also compute
the interacting energy of the $S_z=1$ excitations in the ground state.
The total energy of the system equals
\begin{equation}\label{eAvi}
E = \int d\theta (\Delta\cosh (\theta ) - H)\rho(\theta ).
\end{equation}
If we vary the density of occupied particles and holes, $\rho\rightarrow
\rho+ \delta\rho$ and
$\tilde\rho\rightarrow
\tilde\rho+ \delta\tilde\rho$, the total energy varies accordingly
$$
\delta E = \int d\theta (\Delta\cosh (\theta ) - H)\delta\rho (\theta ).
$$
But we can also express this variation in energy in terms of the 
interacting or dressed
energies of the excitations:
$$
\delta E = \int d\theta 
\big(\delta\rho (\theta ) \ep^+(\theta ) - \delta\tilde\rho (\theta )
\ep^-(\theta )\big).
$$
$\ep^+ (\theta )/\ep^-(\theta )$ mark the energies needed to excite a particle/hole
above the ground state.  At zero temperature they are defined such that
\begin{eqnarray}\label{eAvii}
\ep^+ (\theta ) &=& \Bigg\{{> 0, ~~~~ |\theta| > \theta_F \atop 
= 0, ~~~~ |\theta| < \theta_F} ;\cr\cr
\ep^- (\theta ) &=& \Bigg\{{ = 0, ~~~~ |\theta| > \theta_F \atop
< 0, ~~~~ |\theta| < \theta_F}.
\end{eqnarray}
Together $\ep^{\pm}$ make up a monotonic, smooth function
via $\ep = \ep^+ + \ep^-$.
Comparing the two expressions for $\delta E$ and using the constraint
on the variation in $\rho$ and $\tilde\rho$ coming from (\ref{eAiv}),
$$
\delta\rho + \delta\tilde\rho = \int d\theta' \delta\rho (\theta ') 
\Gamma_{++}(\theta - \theta'),
$$
we arrive at equation (\ref{exxiv}).

\section{Determination of Dressed Scattering Phase}

In this appendix we compute the dressed two body scattering
phase of a particle-hole excitation.  To compute this phase
we examine how the momentum of one of the added excitations is altered
by the addition of both excitations to the system.

The two-body scattering phase of the particle-hole excitation is
defined by
\begin{equation}\label{eBi}
\delta_{ph} (\theta_p ,\theta_h) = L( p_p (\theta_p) - p_{p0} (\theta_p )),
\end{equation}
where $p_p(\theta)$ is the momentum of the particle when the particle/hole
pair is present, while $p_{p0}(\theta)$ is what the momentum of the
particle would be without the hole and without the effect of
the particle's presence on the sea of excitations already present
in the $T=0$ ground state.

Let $\theta_{p},\theta_{h}$ 
be the rapidities of the particle and hole
we intend to add to the system.  Without having added either excitation,
the dressed momentum of the particle is given by
\begin{eqnarray}\label{eBii}
p_{p0} (\theta_p ) &=& \Delta \sinh (\theta_p ) 
+ 2\pi \int^{\theta_F}_{-\theta_F}
d\theta n_0 (\theta ) \phi (\theta_p - \theta);\cr\cr
\phi (\theta ) &=& {1\over 2\pi i}\log S_{++}(\theta ),
\end{eqnarray}
where $n_0\equiv \rho + \tilde\rho$ 
is the density of states given in (\ref{exxii}).  This
is the density of states {\it without} either the particle or hole 
present. Once the particle and hole are added, the density of
the ground-state sea is altered.  
Rather than being given by (\ref{exxii}), it is governed by
\begin{eqnarray}\label{eBiii}
n (\theta ) = {\Delta \over 2\pi} \cosh (\theta ) 
+ {1\over L} (\Gamma_{++} (\theta - \theta_p) - \delta (\theta - \theta_h))
+ \int^{\theta_F}_{-\theta_F} d\theta' n(\theta' ) \Gamma_{++} 
(\theta - \theta').
\end{eqnarray}
The $1/L$ term $(\Gamma (\theta - \theta_p) -\delta
(\theta-\theta_h))$ represents the disturbance the particle-hole
excitation produces in the density of states.  The $\delta (\theta
-\theta_h)$ arises from removing the particle to create the hole,
while $\Gamma (\theta -\theta_p)$ arises from
including the particle in the integral $\int d\theta\,\rho\,\Gamma_{++}$: 
the presence of this particle alters the density of states in the same way
the continuum of particles in the ground state does.

Taking into account the alteration the addition the particle-hole
excitation has upon the density of states, the momentum of the added
particle becomes
\begin{equation}\label{eBiv}
p_{p} (\theta_p ) = \Delta \sinh (\theta_p ) 
+ 2\pi \int^{\theta_F}_{-\theta_F}
d\theta  n (\theta ) \phi (\theta_p - \theta) .
\end{equation}
As $\theta_F$ is unshifted by the addition of the particle-hole pair
we are able to write $n(\theta )$ as
\begin{equation}\label{eBv}
n(\theta ) = n_0 (\theta ) + {1\over L}n_1 (\theta ) - 
{1\over L}\delta (\theta - \theta_h).
\end{equation}
Substitution of (\ref{eBv}) into (\ref{eBiii}) then allows the scattering
phase to be reduced to
\begin{eqnarray}\label{eBvi}
\delta_{ph}(\theta_p ,\theta_h) &=& -2\pi \phi (\theta_p-\theta_h)
+  2\pi \int^{\theta_F}_{-\theta_F}
d\theta  n_1 (\theta ) \phi (\theta_p - \theta );\cr\cr
n_1 (\theta ) &=& \Gamma_{++} (\theta - \theta_p) - 
\Gamma_{++} (\theta - \theta_h)
+ \int^{\theta_F}_{-\theta_F} d\theta' n_1(\theta' ) \Gamma_{++} 
(\theta - \theta').
\end{eqnarray}
We are interested in computing the scattering phase right at
the Fermi surface, i.e. $\delta_{ph}(\theta_p=\theta_F,\theta_h=\theta_F)$.
But we then immediately see
\begin{equation}\label{eBvii}
\delta_{ph}(\theta_F,\theta_F) = -2\pi\phi (0) = -\pi.
\end{equation}
Hence the low-energy particle-hole S-matrix is, 
$S=e^{i\delta_{ph}(\theta_F,\theta_F)} = -1$, as we claimed.  
Here we have focused upon the changes in the momentum of the particle.
If we had examined the momentum of the hole instead we would have arrived at
an identical conclusion.

\section{Reduction of a Ferromagnetic 
Spin-1/2 Ladder to an Effective Spin-1/2 Chain}

In this section we consider a map reducing a spin-1/2 ladder with
ferromagnetic rung interactions in a magnetic field to an effective
spin-1/2 chain.  The essential idea behind the map is already
discussed in a number of articles \cite{approach1,gia}.  However there
the treatment considered a ladder with antiferromagnetic rung interactions.
Here we adapt the map to ladders with ferromagnetic rungs.

\begin{figure}
\centerline{\epsfxsize=2.0in\epsffile{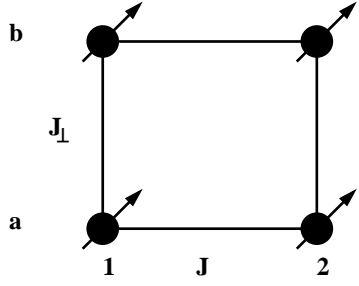}}
\caption{A two rung spin ladder.}
\end{figure}

For the sake of simplicity we consider a two-rung ladder, as pictured
in Figure 8.  The
Hamiltonian of the ladder is given by,
\begin{eqnarray}\label{eCi}
{\cal H} &=& {\cal H}_J + {\cal H}_{J_\perp}\cr\cr
&=&  J(S_{a1}\cdot S_{a2} + S_{b1}\cdot S_{b2}) \cr\cr
&& + J_{\perp\perp}(S^+_{a1}S^-_{b1}+S^+_{a2}S^-_{b2} + {\rm h.c.})
+ J_{\perp z}(S^z_{a1}S^z_{b1}+S^z_{a2}S^z_{b2}) \cr\cr
&& + H(S^z_{a1} + S^z_{b1} + S^z_{a2} + S^z_{b2}).
\end{eqnarray}
We will suppose that $J>0$ while $J_{\perp\perp} < J_{\perp z} < 0$.

We first treat the limit $J=0$ where the rungs decouple.  The eigenstates
on each rung with their corresponding energies are then 
\begin{eqnarray}\label{eCii}
|+\rangle &=& |\uparrow\uparrow\rangle; ~~~~ {J_{\perp z}\over 4} - H\cr\cr
|0\rangle &=& {1\over \sqrt{2}}(|\uparrow\downarrow\rangle 
+ |\downarrow\uparrow\rangle ); ~~~~ 
{J_{\perp\perp}\over 2} - {J_{\perp z}\over 4}\cr\cr
|-\rangle &=& |\downarrow\downarrow\rangle; ~~~~ {J_{\perp z}\over 4} + H\cr\cr
|s\rangle &=& {1\over \sqrt{2}}(|\uparrow\downarrow\rangle 
- |\downarrow\uparrow\rangle ); ~~~~ 
-{J_{\perp\perp}\over 2} - {J_{\perp z}\over 4},
\end{eqnarray}
where the first three states are that of the triplet while the
remaining state correspond to the singlet.
The two states $|+\rangle$ and $|0\rangle$ are of lowest energy.  It
is these two states that we will use in constructing the effective
spin-1/2 chain.

To take into account ${\cal H}_J$ we perform second order perturbation
theory.  We find that the energies of the four basis states
describing the low energy Hilbert space of both rungs, $|++\rangle$,
$1/\sqrt{2}(|+0\rangle \pm |0+\rangle)$, and $|00\rangle$, are shifted
as follows:
\begin{eqnarray}\label{eCiii}
|++\rangle :  && {J\over 2} \cr\cr
1/\sqrt{2}(|+0\rangle \pm |0+\rangle) : && \pm {J\over 2} \cr\cr
|00\rangle : && \alpha \equiv {J^2\over 2(J_{\perp\perp} - J_{\perp z})} 
+ {J^2\over 8J_{\perp\perp}}.
\end{eqnarray}
Understanding $|+\rangle = |\tilde\uparrow\rangle$ and $|-\rangle 
= |\tilde\downarrow\rangle$, we can write down an effective
system of interaction spin-1/2's:
\begin{equation}\label{eCiv}
{\cal H}_{eff} = 
{J^e_\perp \over 2} (\sigma^+_1\sigma^-_2 + \sigma^+_2\sigma^-_1)
+ J^e_z \sigma^z_1\sigma^z_2 - H^e(\sigma^z_1 + \sigma^z_2) + c,
\end{equation}
where the various effective couplings are given in terms of the original
couplings by
\begin{eqnarray}\label{eCv}
J^e_\perp &=& J ;\cr\cr
J^e_z &=& J/2 + \alpha ;\cr\cr
H^e &=& {H\over 2} + {\alpha\over 4} + {1\over 4}(J_{\perp\perp} - J_{\perp z}
- {J \over 2}) ;\cr\cr
c &=& {J_{\perp\perp}\over 2} - H + {J\over 8} + {\alpha \over 4}.
\end{eqnarray}
We point out that as $\alpha$ can be negative, so can $J^e_z$.
$J^e_z$ negative is essential to correctly describing the physics
of the spin-1 chain.

Having reduced the spin ladder with ferromagnetic rung couplings to
a spin-1/2 chain we now bosonize.  Applying the standard results
(see \cite{affrev}) we find that the associated Lagrangian is
\begin{eqnarray}\label{eCvi}
{\cal L} &=& {K\over 2\pi}\partial^\mu\Phi\partial_\mu\Phi;\cr\cr
K &=& {1 \over (1 + {2J^e_z\over \pi J^e})},
\end{eqnarray}
while the fields are given by
\begin{eqnarray}\label{eCvii}
n^-(x) &=& {1\over \sqrt{2}}(S^-_a(x) + S^-_b(x)) = \sigma^- (x) = 
e^{-i\Phi (x)}(1+ c_1\cos(2\pi (\Theta (x )-Mx)));\cr\cr
M_z (x) &=& S^z_{a}(x) + S^z_{b}(x) = \sigma^z (x) + 1/2 = 
-{K\over v_F\pi}\del_\tau \Phi (x) + c \cos (2\pi(\Theta (x)-Mx)) .
\end{eqnarray}
Here $M$ is the magnetization density of the spin-1/2 ladder (and so
the corresponding spin-1 chain).  In terms of the magnetization, $M^e$,
of the effective spin-1/2 chain, $M = M^e+1/2$.
We see $n^-$ is defined in terms of the $k_y =0$ component of the ladder
spin operators.  Here we thus expect $n^-$ to encode power law correlations
for wavevectors near $k=(\pi , 0)$.  In a ladder with antiferromagnetic
rungs, power law correlations would be seen instead for $k=(\pi ,\pi)$.

With $J^e_z < 0$ (provided $\alpha$ is sufficiently large) we see
that $K > 1$ in correspondence with both our analysis and the results
of \cite{approach2}.  This marks out the main difference between ladders
with ferromagnetic rungs and ladders with antiferromagnetic rungs.
With antiferromagnetic rungs, a similar analysis leads to $J_z > 0$
and so $K < 1$ (\cite{approach1,gia}).
\vfill\eject

\end{document}